\documentclass[11pt,twoside]{article}

\usepackage{asp2004}
\usepackage{epsf}
\usepackage{psfig}
\usepackage{lscape}

\begin{document}
\title{Measuring High Moments of Galaxy Angular Distribution\\
in the First Look Survey using Infrared Array Camera}   
\author{Fan Fang and the FLS Team}   
\affil{Spitzer Science Center, 314-6, California Institute of Technology, Pasadena, CA 91125}    

\begin{abstract} 
We present the results of using three methods, the probability distribution
function, the R\'{e}nyi information, and the multi-fractals, to measure the
high moments of galaxy angular distribution in the Infrared Array Camera
data of the Spitzer First Look Survey.  This is the first time that
R\'{e}nyi information is explicitly used to describe the large-scale galaxy
spatial distribution, and our galaxy samples are also the first of the kind
in their wavelengths and sensitivities.
\end{abstract}

\begin{figure}[!ht]
\vspace{-0.2in}
\plottwo{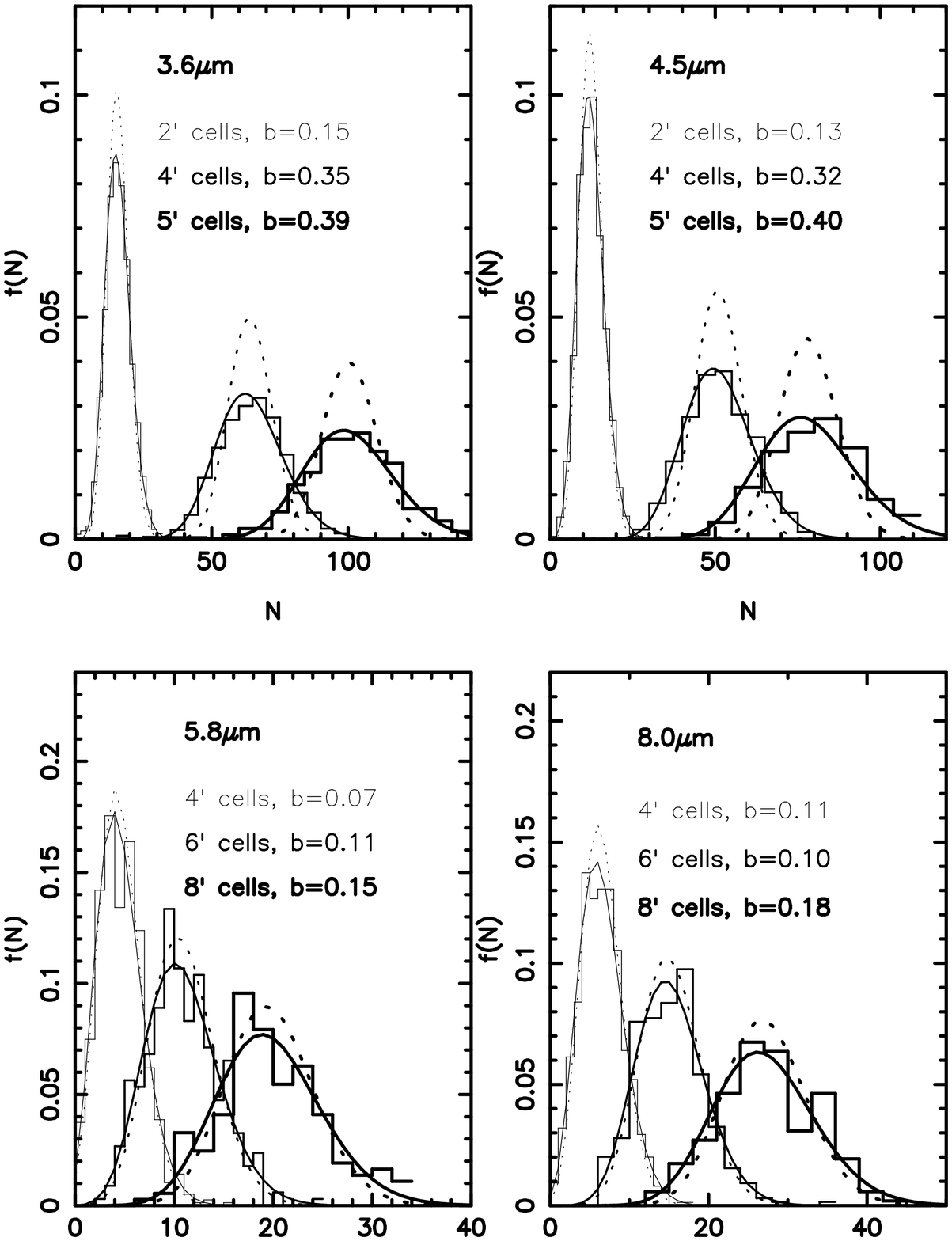}{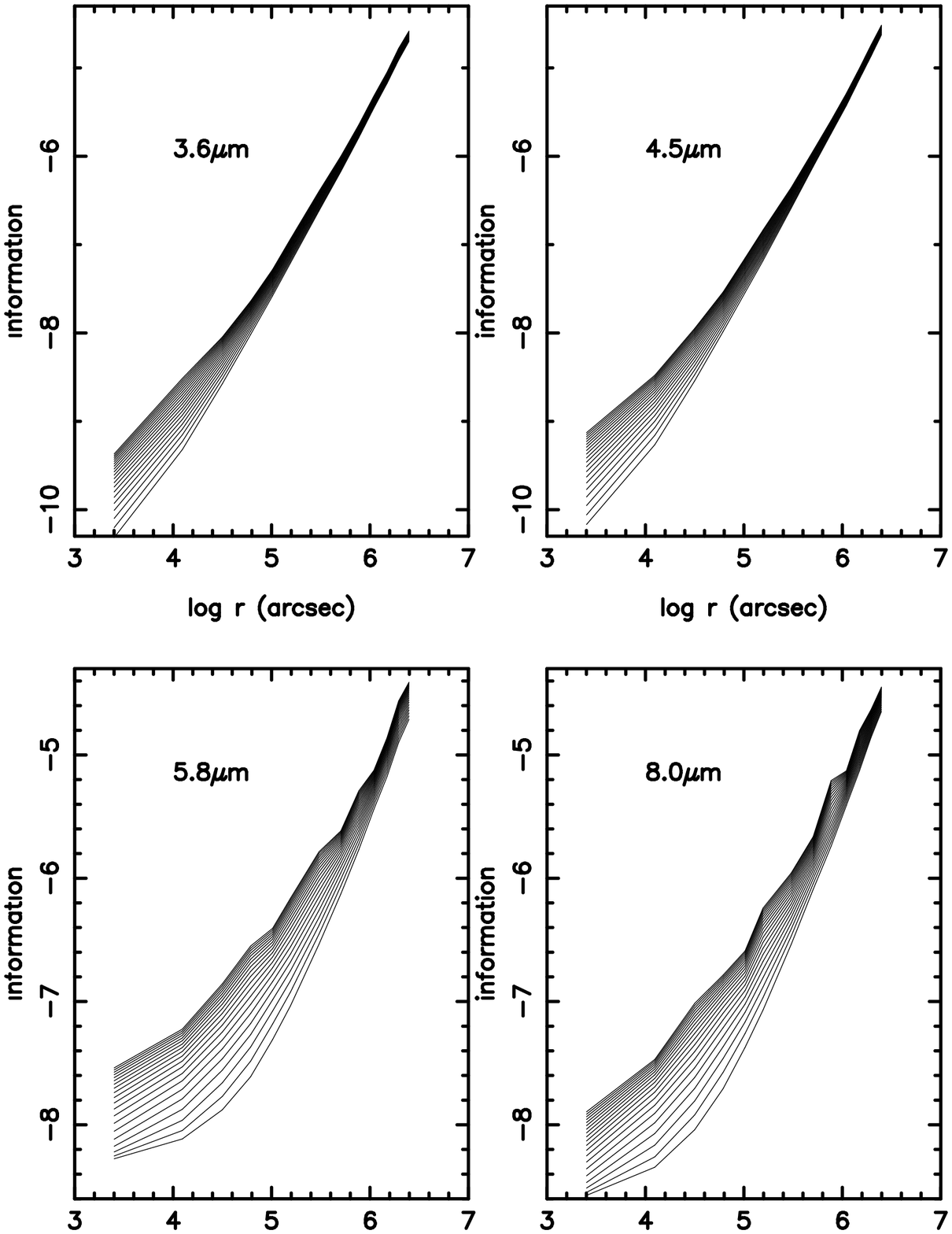}
\end{figure}

A previous study \citep{fang04} measured the 2-point angular correlation
functions for galaxy samples from all four wavelengths of the Infrared
Array Camera (IRAC) in the Spitzer First Look Survey (FLS).  Measuring
the second moment, the 2-point statistics do not completely describe a
non-Gaussian spatial density field.  Here we extend the study to measuring
high moments for the same samples, using the probability distribution
function (PDF), the R\'{e}nyi information, and the multifractal descriptions.
For all measurements the two-dimensional area covered by the IRAC samples
is divided into contiguous square cells of varying sizes.  We discard all
cells that contain invalid mosaic pixels determined by masks that used
to establish the IRAC samples.  The boundary effect and selection bias
are at minimum.

The PDF contains all the high moments of the galaxy spatial distribution.
It is calculated by counting the number of galaxies in the cells and
establishing normalized histograms.  In the left figure of the previous
page we show our PDF results.  For each sample it is measured at 3 different
scales.  For each measurement we plot the fit of the theoretical Gravitational
Quasi-equilibrium Distribution Function \citep{sh84} (solid lines) and the
Poisson distribution of the same average galaxy count for comparisons.
The deviation of the PDF from Poisson caused by galaxy clustering is
obvious at large scales.  The gravitational distribution function fits well
at all scales.  The value of the fitting parameter $b$ shows the strength of
clustering.  Theoretically it is the ratio of the correlation potential energy
and twice the kinetic energy.

For the first time, we explicitly apply the R\'{e}nyi Information
to describe the spatial distribution of the galaxies.
The $n$-order R\'{e}nyi Information is defined by \citep{renyi70}
$I_{n} = \frac{1}{n-1}\log \Sigma p^{n}$,
where $p$ is the probability of finding a galaxy in a cell.  Since $p$
is described by the PDF, the $n$-order R\'{e}nyi Information directly
measures the $n$-moment of the distribution.  In the right figure of the
previous page we plot the R\'{e}nyi information measurements, in "bits"
of unit log2, over a range of cell sizes placed on the IRAC samples, and from
$n = 1$ up to $20$ (bottom to top).  The crowding of the lines at
high $n$ shows that the high moments are constrained in a narrow
information space, a desired property for studying the moments of
arbitrary high orders.

The rate of information change with cell sizes gives an estimate of the
fractal dimension of galaxy distribution.  The information figure
indicates that galaxies occupy space compactly at small scales, and the
distribution becomes more uniform at large scales.  The multi-fractal
dimensions measured across increasing information orders decrease and approach
a limit.  This is illustrated above in the left figure.  The top right
figure shows the spectra of the multi-fractal scaling index,
related to multi-fractal dimensions by a Legendre transformation.  As the
spectra goes to zero, the scaling index measures the finite value of
the multi-fractal dimension at infinite information order.  Therefore
a multi-fractal simply describes the hierarchy of the moments in
galaxy spatial distribution.

\begin{figure}
\plotfiddle{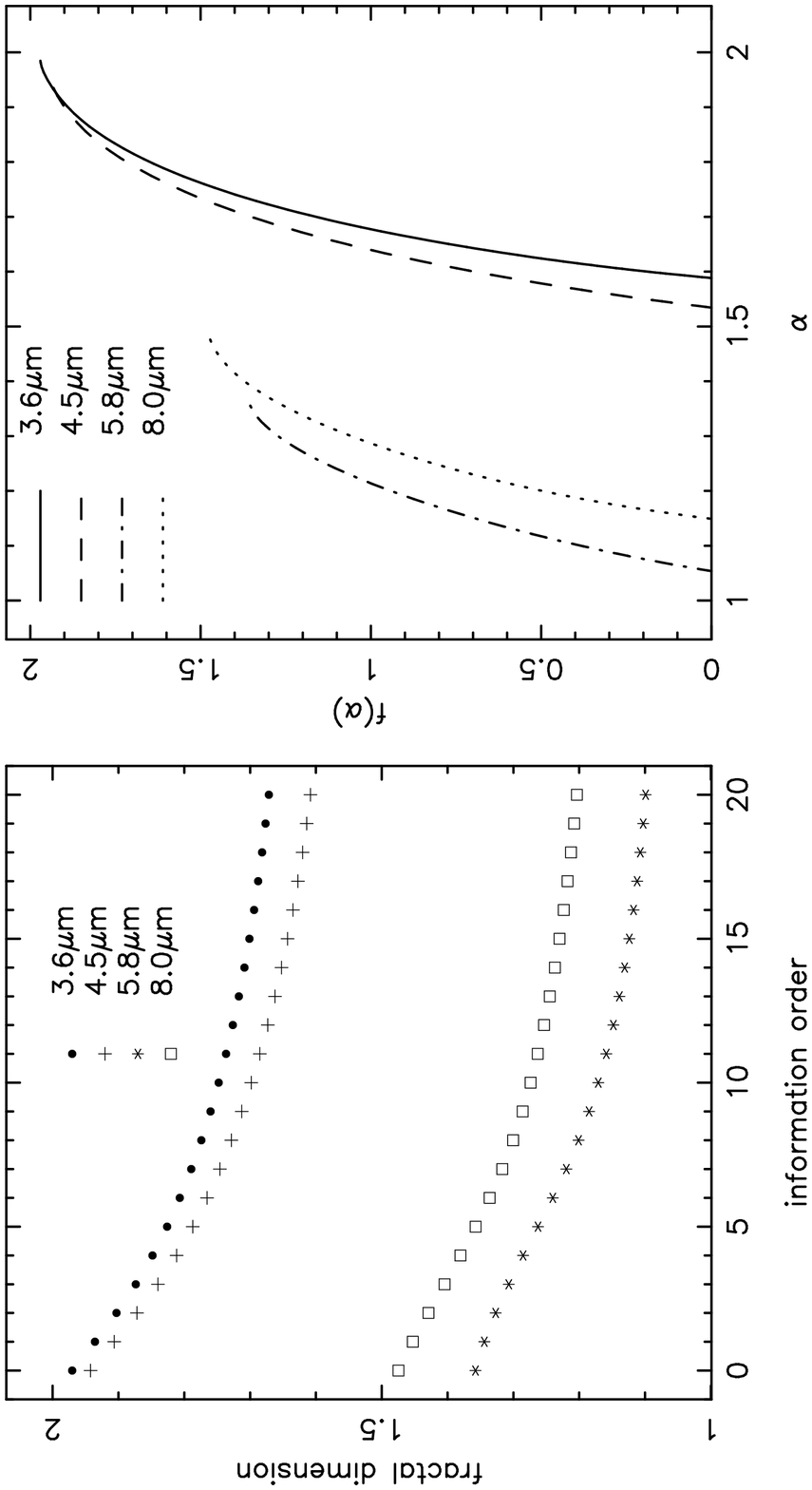}{1.0in}{270}{30}{30}{-200}{136}
\end{figure}

\acknowledgements 

Support for this work was provided by NASA through the Jet Propulsion
Laboratory, California Institute of Technology under NASA contract 1407.

\end{document}